\documentclass[journal]{IEEEtran}

\usepackage{tikz}
\usepackage{animate}
\usepackage{booktabs}

\graphicspath{{figures/}}
\newcommand\figwidth{8.9} 

\usepackage[colorlinks]{hyperref}
\hypersetup{
	colorlinks=true, 
	linkcolor=blue!70!black, 
	citecolor=red!70!black, 
	filecolor=blue!70!black, 
	urlcolor=blue!70!black, 
}

\RequirePackage{xcolor} 

\definecolor{PairedA}{RGB}{166, 206, 227}
\definecolor{PairedB}{RGB}{ 31, 120, 180}
\definecolor{PairedC}{RGB}{178, 223, 138} 
\definecolor{PairedD}{RGB}{ 41, 128,  35}   
\definecolor{PairedE}{RGB}{251, 154, 153}
\definecolor{PairedF}{RGB}{182,  21,  22}   
\definecolor{PairedG}{RGB}{253, 191, 111}
\definecolor{PairedH}{RGB}{255, 127,   0}
\definecolor{PairedI}{RGB}{202, 178, 214}
\definecolor{PairedJ}{RGB}{106,  61, 154}
\definecolor{PairedK}{RGB}{255, 255, 153}
\definecolor{PairedL}{RGB}{177,  89,  40}

\RequirePackage[cmex10,intlimits]{amsmath}
\RequirePackage{amsfonts}
\RequirePackage{amssymb}

\providecommand{\Quot}[1]{``{#1}"}              
\providecommand{\V}[1]{\boldsymbol{#1}}         
\providecommand{\M}[1]{\mathbf{#1}}             
\providecommand{\T}[1]{\mathrm{#1}}             
\providecommand{\OP}[1]{{\mathcal{#1}}}         

\providecommand{\herm}{\mathrm{H}} 
\providecommand{\trans}{\mathrm{T}}

\providecommand{\srcRegion}{\ensuremath{\varOmega}} 
\providecommand{\basisFcn}{\V{\psi}}

\providecommand{\Ivec}{\ensuremath{\M{I}}}

\providecommand{\yvec}{\ensuremath{\M{y}}}

\providecommand{\Zmat}{\ensuremath{\M{Z}}}

\providecommand{\Ymat}{\ensuremath{\M{Y}}}

\providecommand{\Cmat}{\ensuremath{\M{C}}}

\newcommand{\ie}{\textit{i.e.}{}}

\newcommand{\cf}{\textit{cf.}{}}

\RequirePackage[nolist]{acronym}
\newacro{MoM}[MoM]{method of moments}
\newacro{MOO}[MOO]{multiobjective optimization}
\newacro{CM}[CM]{characteristic mode}
\newacro{PEC}[PEC]{perfect electric conductor}
\newacro{ESA}[ESA]{electrically small antenna}
\newacro{PMC}[PMC]{perfect magnetic conductor}
\newacro{EP}[EP]{eigenvalue problem}
\newacro{GEP}[GEP]{generalized eigenvalue problem}
\newacro{EFIE}[EFIE]{electric field integral equation}
\newacro{SVD}[SVD]{singular value decomposition}
\newacro{RWG}[RWG]{Rao-Wilton-Glisson}
\newacro{EM}[EM]{electromagnetic}
\newacro{dof}[d-o-f]{\mbox{degrees-of-freedom}}

\begin{document}
\title{Inversion-Free Evaluation of Nearest Neighbors in Method of Moments}
\author{Miloslav~Capek,~\IEEEmembership{Senior~Member,~IEEE,}
        Lukas~Jelinek,
        and~Mats~Gustafsson,~\IEEEmembership{Senior~Member,~IEEE}
\thanks{Manuscript received \today; revised \today.}
\thanks{This work was supported by the Czech Science Foundation under project~\mbox{No.~19-06049S}. The work of M.~Capek was supported by the Ministry of Education, Youth and Sports through the project CZ.02.2.69/0.0/0.0/16\_027/0008465.}
\thanks{M.~Capek and L.~Jelinek are with the Department of Electromagnetic Field, Faculty of Electrical Engineering, Czech Technical University in Prague, 166~27 Prague, Czech Republic (e-mail: \mbox{miloslav.capek@fel.cvut.cz}, \mbox{lukas.jelinek@fel.cvut.cz}).}
\thanks{M.~Gustafsson is with the Department of Electrical and Information Technology, Lund University, 221~00 Lund, Sweden (e-mail: mats.gustafsson@eit.lth.se).}
}

\maketitle

\begin{abstract}
A recently introduced technique of topology sensitivity in method of moments is extended by the possibility of adding degrees-of-freedom (reconstruct) into underlying structure. The algebraic formulation is inversion-free, suitable for parallelization and scales favorably with the number of unknowns. The reconstruction completes the nearest neighbors procedure for an evaluation of the smallest shape perturbation. The performance of the method is studied with a greedy search over a Hamming graph representing the structure in which initial positions are chosen from a random set. The method is shown to be an effective data mining tool for machine learning-related applications.
\end{abstract}


\IEEEpeerreviewmaketitle
\section{Introduction}
\label{sec:intro}

\IEEEPARstart{S}{hape} synthesis, a technique of constructing a particular body from piece-wise constant materials, is an unsolved problem across many engineering branches. The major obstacle is a $2^N$~combinatorial explosion~\cite{Lawler_CombinatorialOptimization} for $N$~\ac{dof} that arises from its binary nature: each unknown is associated with a given material or with vacuum~\cite{RahmatMichielssen_ElectromagneticOptimizationByGenetirAlgorithms}. While the formulation can be relaxed by introducing continuous variables, as in the case of topology optimization~\cite{BendsoeSigmund_TopologyOptimization}, the solution is finally rounded with respect to a given threshold~\cite{2016_Liu_AMS}. This last step is encumbered with difficulties such as non-uniqueness and instability~\cite{Deschamps+Cabayan1972}.

Contemporary solutions to shape optimization are mostly parametric sweeps, heuristic algorithms~\cite{Simon_EvolutionaryOptimizationAlgorithms}, and, recently, machine learning~\cite{Goodfellow_DeepLearning}. All these techniques share a common feature: a demand on vast amounts of samples for which a fitness function has to be evaluated. Therefore, large datasets are dealt with during shape optimization.

In this contribution, the binary nature of the optimization problem is kept in its original form accepting NP complexity, and a novel method of topology sensitivity~\cite{Capeketal_ShapeSynthesisBasedOnTopologySensitivity} is adopted and extended by the possibility of reconstructing the structure. The resulting local algorithm is based on an investigation of nearest neighbors in a Hamming graph~$H(N,2)$~\cite{VanDam_etal-DistanceRegularGraphs}. It utilizes \ac{MoM}~\cite{Harrington_FieldComputationByMoM} and the Sherman-Morrison-Woodbury identity~\cite{GolubVanLoan_MatrixComputations, 1989_Hager_SIAM_R}, a popular scheme for evaluating consequences of local geometry perturbations~\cite{Capeketal_ShapeSynthesisBasedOnTopologySensitivity,1989_Kastner_OnMatrixPartitioning, 1989_Kastner_AnAddOnMethodForTheAnalysisOfScattering, 2010_Laviada_etal-AnalysisOfPartialModificationsOnElLargeBodiesViaCBF, 2018_ChenEtAl_AnalysisOfPartialGeometryModificationProblems}.

Exploration of all shapes with a unit Hamming distance is performed with an inversion-free evaluation. The proposed method is fine-grained (with an expected linear speed-up when parallelization is used), well-suited for vectorization, and its implementation has the potential to evaluate millions of mid-size antenna candidates per minute on a laptop. As such, it represents an ideal data mining tool for machine learning algorithms~\cite{Liu_etal_AnEfficientMethodForAntennaDesignOptimizationBasedOnECandML} or an apt candidate for a local step in global optimization~\cite{Gregory_etal_FastOptimizationOfEMDesignProblemsUsingCovarianceMatrixAdaptionEvolutionaryStrategy}. Specifically, the training phase of supervised learning~\cite{Chen_AHybridAlgotihmOfDEandMLforEMstructureOptimization} can be significantly shortened. Utilizing the linear regression models~\cite{Chen_AHybridAlgotihmOfDEandMLforEMstructureOptimization}, the proposed technique can provide additional information about first-order perturbations. These claims are supported by a Monte Carlo simulation with a greedy search over the nearest neighbors.

The letter is organized as follows. Topology sensitivity technique is briefly reviewed in Section~\ref{sec:perturbInMoM} and extended by the possibility of reconstructing a previously reduced shape. Section~\ref{sec:nearNeigh} shows that an iterative evaluation of all nearest neighbours of an actual shape can be employed in a greedy algorithm. Performance of the greedy algorithm based on the nearest neighbours search is statistically studied in Section~\ref{sec:monteCarlo}. The letter concludes in Section~\ref{sec:concl}.

\section{Effective Shape Perturbation in MoM}
\label{sec:perturbInMoM}

\begin{figure}
\centering
\includegraphics[width=\figwidth cm]{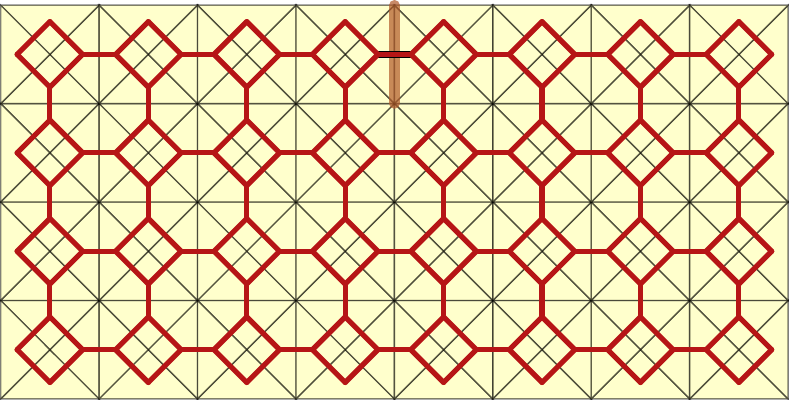}
\caption{A discretized rectangular plate with a side aspect ratio of 2:1. Binary optimization is defined over a graph depicted by red lines representing a connectivity of basis functions. The thick vertical line represents the position of a delta gap source~\cite{Balanis1989}.}
\label{fig:fig1C}
\end{figure}

\ac{MoM}~\cite{Harrington_FieldComputationByMoM} for a fixed discretization and a set of piece-wise basis functions~$\left\{\basisFcn_n\left(\V{r}\right)\right\}$ is considered here as the starting point, see Fig.~\ref{fig:fig1C}. The presence of basis functions is a subject of binary optimization, see \cite{Capeketal_ShapeSynthesisBasedOnTopologySensitivity} for details. The primary quantities being operated on are the impedance matrix ~$\Zmat \in \mathbb{C}^{N\times N}$~\cite{Harrington_FieldComputationByMoM}, which carries information about the electromagnetic behaviour of an actual shape, and a set of matrix operators $\left\{\M{A}_m \right\}$ representing criterion function~$p$ in a form
\begin{equation}
\label{eq:fitFunction}
p \left(\Ivec\right) = h \left( 
\left[ \begin{array}{*{30}{c}}
	\cdots , & \Ivec^\herm \M{A}_m \Ivec \, , & \cdots
\end{array} \right] \right),
\end{equation}
where~$\Ivec$ is a column vector of current expansion coefficients~\cite{Harrington_FieldComputationByMoM}, and $h$ is an arbitrary function operating over individual quantities $\Ivec^\herm \M{A}_m \Ivec$, $m \in \left\{1,\dots,M\right\}$. Let us further assume that the~$f$th basis function is fed where index~$f$ is fixed.

\begin{figure}
\centering
\includegraphics[width=\figwidth cm]{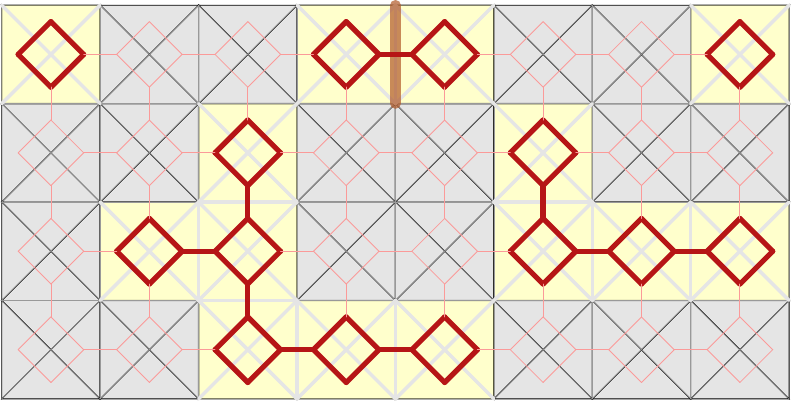}
\caption{Randomly generated initial structure~$\srcRegion_\OP{E}$. Gray and yellow pixels represent vacuum and metal, respectively. All basis functions fully residing in the conducting region are enabled, the others are disabled. Position of the feeder is highlighted by the thick vertical line and is always a part of the metallic motif. All disabled \ac{dof} can be subject of addition, \ie{}, they belong to set~$\OP{B}_+$, and enabled \ac{dof} can be subject of removal, \ie{}, they belong to set~$\OP{B}_- = \OP{E} \setminus \left\{ f\right\}$.}
\label{fig:fig1D}
\end{figure}

Topology sensitivity is defined as~\cite{Capeketal_ShapeSynthesisBasedOnTopologySensitivity}
\begin{equation}
\label{eq:topoSens}
\tau_b \left( p, \srcRegion_\OP{E} \right) = - \Big( p \left( \Ivec_\OP{E} \right) - p \left( \Ivec_b \right) \Big),
\end{equation}
expressing a difference between the performance of the actual shape~$\srcRegion_\OP{E}$ and of its smallest perturbations realized via individual removals or additions of the $b$th basis function, $b \in \mathcal{B}$. The symbol~$\OP{E}$~denotes a set of all \Quot{enabled} \ac{dof} and~$\OP{B}$~denotes a set of all investigated edges, $\OP{B} = \left\{\OP{B}_-, \OP{B}_+ \right\}$, where $\OP{B}_-$~is a set of \ac{dof} to be removed (one by one) and $\OP{B}_+$~is a set of \ac{dof} to be added (one by one), see Fig.~\ref{fig:fig1D} for a particular example.

\subsection{Removal of Basis Functions}

Shape reduction via the removal of \ac{dof} with indices~$b \in \OP{B}_-$ is possible with a linear asymptotic complexity as
\begin{equation}
\label{eq:topoRemI}
\left[\Ivec_{\OP{E} \setminus \OP{B}_-} \right] = 
\left[ {\begin{array}{*{40}{c}} \cdots , & \yvec_{\OP{E},f} - \displaystyle\frac{Y_{fb}}{Y_{bb}} \yvec_{\OP{E},b} \, , & \cdots \end{array}} \right]
l_f V_0,
\end{equation}
with an effective evaluation of \eqref{eq:fitFunction} as 
\begin{equation}
\label{eq:quadFormEval1}
\M{p}_{\OP{E} \setminus \OP{B}_-} =  h \left( \OP{D} \left(\M{A}_{\OP{E},m}, \left[\Ivec_{\OP{E} \setminus \OP{B}_-} \right]\right) \right),
\end{equation}
where
\begin{equation}
\label{eq:quadFormEval2}
\OP{D} \left(\M{A}_{\OP{E}}, \left[\Ivec_{\OP{E} \setminus \OP{B}_-} \right]\right) = \T{diag}\left(\left[\Ivec_{\OP{E} \setminus \OP{B}_-} \right]^\herm \M{A}_{\OP{E}} \left[\Ivec_{\OP{E} \setminus \OP{B}_-} \right]\right).
\end{equation}
The symbols introduced in \eqref{eq:topoRemI}--\eqref{eq:quadFormEval2} read:~$\yvec_{\OP{E},f}$ and~$\yvec_{\OP{E},b}$ are the~$f$th and ~$b$th columns of the admittance matrix~$\Ymat_\OP{E} = \Zmat_\OP{E}^{-1}$, respectively,~$Y_{fb}$ and~$Y_{bb}$ are admittance matrix elements, $l_f$ is the length of the fed edge, and $V_0 = 1$\,V. Matrix operators~$\left\{\M{A}_{\OP{E},n}\right\}$ have only lines and columns corresponding to the entries in set~$\OP{E}$ {and matrices in a form of~$\left[ \Ivec \right]$ contain columns of current expansion coefficients corresponding to the investigated perturbations.

The edge corresponding to the worst topology sensitivity is selected (here denoted by index~$b$) and removed (disabled) by virtue of an admittance matrix update
\begin{equation}
\label{eq:topoRemMatUpdate}
\Ymat_{\OP{E} \setminus b} = \Cmat_{\OP{E} \setminus b}^\trans \left( \Ymat_\OP{E} - \left(\frac{\yvec_{\OP{E},b}}{Y_{bb}} \right) \yvec_{\OP{E},b}^\trans \right) \Cmat_{\OP{E} \setminus b},
\end{equation}
where $\Cmat_{\OP{E} \setminus b}$ is a permutation matrix,
\begin{equation}
\label{eq:topoRemCmat}
C_{\OP{E} \setminus b, nn} = \left\{
\begin{array}{lll}
0 & \Leftrightarrow & n = b, \\
1 & \Leftrightarrow & \mathrm{otherwise}, \\
\end{array}
\right.
\end{equation}
in which all zero columns are removed. A removal of one basis function thus means reduction of the admittance matrix dimension by one.

\subsection{Addition of Basis Functions}

The basis function removal technique~\eqref{eq:topoRemI} is excellent for investigating topological sensitivity \eqref{eq:topoSens}, however, a nonexistence of a technique of a basis function addition commonly caused a premature deadlock~\cite{Capeketal_ShapeSynthesisBasedOnTopologySensitivity}. Here, the addition of a basis function is introduced by further applying the Sherman-Morrison-Woodbury identity~\cite{GolubVanLoan_MatrixComputations} which results in
\begin{equation}
\label{eq:topoAddI}
\begin{split}
& \left[ \Ivec_{\OP{E} \cup \OP{B}_+} \right] = \\
& \quad \left[ \begin{array}{*{30}{c}}
\cdots, & \Cmat_{\OP{E} \cup b}^\trans \left(
\left[ \begin{array}{*{10}{c}}
	\yvec_f \\
	0
\end{array} \right]
+
\dfrac{x_{fb}}{z_b}
\left[ \begin{array}{*{10}{c}}
	\M{x}_b \\
	-1
\end{array} \right] \right) \, , & \cdots \\
	\end{array} \right] l_f V_0,
\end{split}
\end{equation}
where
\begin{equation}
\label{eq:topoAuxVars}
\M{x}_b = \Ymat \M{z}_b, \quad z_b = Z_{bb} - \M{z}_b^\trans \M{x}_b,
\end{equation}
and permutation matrix~$\Cmat_{\OP{E} \cup b}$ provides a correct ordering of the basis functions since the basic Sherman-Morrison-Woodbury identity demands that the  basis function added must be the last one. Entries of $\Cmat_{\OP{E} \cup b}$ read
\begin{equation}
\label{eq:topoAddCmat}
\Cmat_{\OP{E} \cup b, mn} = \left\{
\begin{array}{lll}
1 & \Leftrightarrow & n = S\left(m\right), \\
0 & \Leftrightarrow & \mathrm{otherwise}, \\
\end{array}
\right.
\end{equation}
with $m \in \left\{ 1, 2, \dots, E+1\right\}$ and where $S$ is a set of target indices if a set~$\left\{ \OP{E}, b\right\}$ is sorted in ascending order.

After deciding which basis function should be added (enabled), an admittance matrix update is performed as follows
\begin{equation}
\label{eq:topoAddMatUpdate}
\Ymat_{\OP{E}\cup b}
=
\frac{1}{z_b} \Cmat_{\OP{E} \cup b}^\trans \left[ {\begin{array}{cc}
	z_b \Ymat + \M{x}_b \M{x}_b^\trans & \displaystyle -\M{x}_b \\
	-\M{x}_b^\trans & \displaystyle 1 \\
	\end{array} } \right] \Cmat_{\OP{E} \cup b},
\end{equation}
with the auxiliary variables defined in~\eqref{eq:topoAuxVars}.

\section{Local Shape Perturbation}
\label{sec:nearNeigh}

Thanks to \eqref{eq:topoRemI}--\eqref{eq:topoRemMatUpdate}, and \eqref{eq:topoAddI}--\eqref{eq:topoAddMatUpdate} the initial shape~$\srcRegion_\OP{E}$ can either be extended or reduced according to its actual topology sensitivity~\eqref{eq:topoSens} to a given parameter~$p$. In order to proceed further, let us represent any properly discretized shape as a binary genus $\M{g} = \left[ b_1, \,\, \cdots \,\, , b_N \right]$ with logical values $b_n \in \left\{0,1\right\}$, where $b_n = 1$ if $n \in \OP{E}$ and $b_n = 0$ otherwise. The Hamming graph~$H(2,N)$ is defined over genes $\M{g}$ in which nearest neighbors can be found and evaluated using~\eqref{eq:topoRemI} and \eqref{eq:topoAddI}, see Fig.~\ref{fig:fig2N} for a sketch of the procedure for $N = 4$. For an arbitrary starting position, a graph in Fig.~\ref{fig:fig2N} can be explored for a locally optimal shape using a greedy algorithm~\cite{Nemhauser_etal_IntegerAndCombinatorialOptimization} following the steepest descent of~\eqref{eq:topoSens}.

\begin{figure}
\centering
\includegraphics[width=7cm]{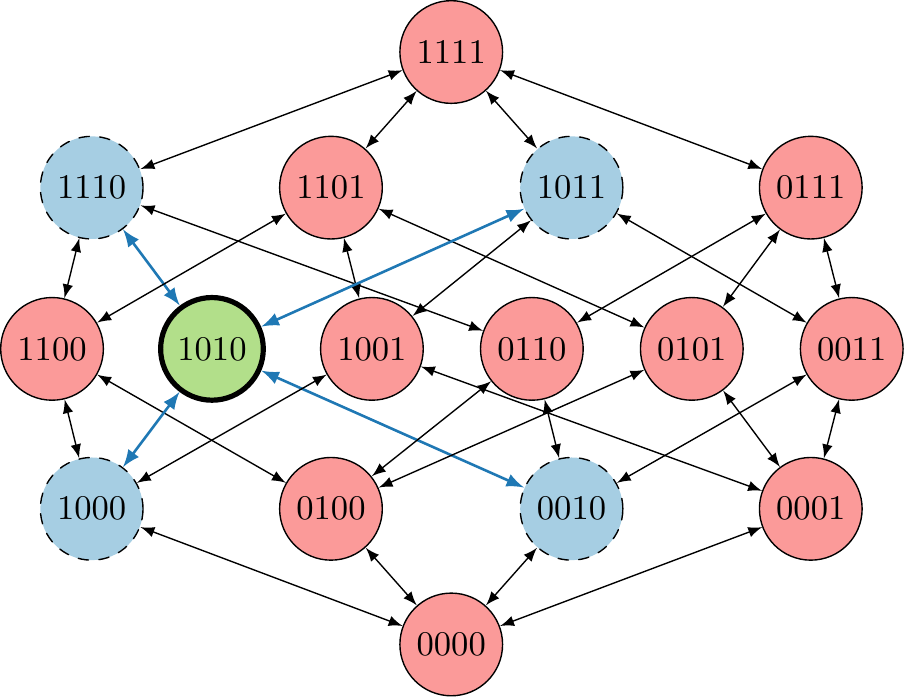}
\caption{Solution space consisting of four \ac{dof} represented as an hierarchic Hamming graph. A Monte Carlo algorithm selected node 1010 (depicted in green color) as the starting point. All the nearest neighbors (in blue) are evaluated effectively without requiring an impedance matrix inversion. They represent the smallest perturbation of the structure, \ie{}, edge addition (upward) or edge removal (downward) evaluated via~\eqref{eq:topoRemI} or~\eqref{eq:topoAddI}, respectively. Movement to the next node is performed in a greedy sense with an admittance matrix update evaluated as~\eqref{eq:topoRemMatUpdate} or~\eqref{eq:topoAddMatUpdate}, respectively. At this point, the procedure starts again.}
\label{fig:fig2N}
\end{figure}

A particular result of the greedy algorithm, based on nearest neighbors, is presented in Fig.~\ref{fig:fig1E} for the starting position depicted in Fig.~\ref{fig:fig1D} and a minimization of a radiation Q-factor which is a parameter of primary importance for electrically small antenna,~\cite{VolakisChenFujimoto_SmallAntennas}. Electrical size is~$ka = 1/2$, with~$k$ being the free-space wavenumber and $a$ being the radius of the sphere fully circumscribing the rectangular region. Actual performance in Q-factor, evaluated according to~\cite{CapekGustafssonSchab_MinimizationOfAntennaQualityFactor}, is normalized throughout the paper with respect to its lower bound being restricted to TM modes only, $Q_\T{lb}^\T{TM}$, see~\cite{Capek2018} and references therein.
As compared to Fig.~\ref{fig:fig1D}, the performance of a shape in Fig.~\ref{fig:fig1E} was improved from $Q_\T{init} / Q_\T{lb}^\T{TM} = 119$ to $Q_\T{final} / Q_\T{lb}^\T{TM} = 1.23$ in $84$ steps. The difference between the initial genome $\M{g}_\T{init}$ and the final genome $\M{g}_\T{final}$ is significant, \cf{} Figs.~\ref{fig:fig1D} and \ref{fig:fig1E}, and is quantified with a normalized Hamming distance between the corresponding genes, \ie{},
\begin{equation}
\label{eq:similarCoef}
\kappa = \frac{1}{N-1} \left\| \M{g}_\T{init} \oplus \M{g}_\T{final} \right\|
\end{equation}
for one basis function being excited by a delta gap. The coefficient~$\kappa = 1$ means the structure was completely changed, while~$\kappa = 0$ means that the structure was not modified at all. In the final case depicted in Fig.~\ref{fig:fig1E}, the~$\kappa$ coefficient equals~$0.47$, which means that approximately one half of the basis functions have been changed when compared with the initial structures.

\begin{figure}[t]
	\begin{center}
		\animategraphics[width=9cm, loop, controls, autoplay, buttonsize=0.75em]{1}{grid4x8topoAnimateQmin_MonteCarlo/grid_4x8_MonteCarlo}{0}{84}
		\caption{Resulting structure from a greedy algorithm evaluating nearest neighbors and following the steepest gradient. It is seen from a comparison with Fig.~\ref{fig:fig1D} that some \ac{dof} were added, some others were removed. The final structure was found in $84$~steps, lowering the fitness function from $Q_\T{init}/Q_\T{lb}^\T{TM} = 119$ for a structure in Fig.~\ref{fig:fig1D} to $Q_\T{final}/Q_\T{lb}^\T{TM} = 1.23$. In total, $15215$~antenna candidates were evaluated in $0.77$~second.}
		\label{fig:fig1E}
	\end{center}
\end{figure}

\section{Monte Carlo Analysis}
\label{sec:monteCarlo}

This section provides a detailed study of the algorithm proposed in the previous section. To this point, multiple runs with random starting positions (feeding position being fixed) were performed and statistically evaluated. As in the previous section, the performance of the algorithm is investigated using Q-factor minimization and a rectangular bounding box with a side aspect ratio of 2:1 with electrical size $ka = 1/2$. Two discretization schemes, $4\times 8$ grid ($N = 180$) and $6\times 12$ grid ($N = 414$), are used as can be seen in the insets in Fig.~\ref{fig:fig3NB} and Fig.~\ref{fig:fig4N}, respectively. The entire procedure was implemented in MATLAB~\cite{matlab}. The matrix operators were evaluated in AToM~\cite{atom}, and all calculations ran on the computer specified in Table~\ref{Tab:compTime1}. The only run-time variables are $\M{g}_\OP{E}$, $\Ymat_\OP{E}$, $\Zmat$ (for shape reconstruction), and~$\left\{ \M{A}_m \right\}$ matrices. In total, $5 \cdot 10^4$ trials were performed for both grids and the overall performance is summarized in Table~\ref{Tab:compTime1} and in Figs.~\ref{fig:fig3NB}--\ref{fig:fig5N}.

\begin{table}[t] 
\centering
\caption{Comparison of shape optimization performance depending on the number of \ac{dof}. The table is extended by a column with the results for an $8\times 16$ grid.}
\begin{tabular}{cccc} 
plate & $4 \times 8$ & $6 \times 12$ & $8 \times 16$ \\[1pt] \toprule
\ac{dof}, $N$ & $180$ & $414$ & $744$ \\[1pt]
runs, $I$ & $5\cdot 10^4$ & $5\cdot 10^4$ & $1\cdot 10^3$ \\[1pt]
comp. time, $T$ [s] & $2.4\cdot 10^3$ & $5.8\cdot 10^4$ & $1.2 \cdot 10^4$ \\[1pt]
evaluated shapes & $7.2\cdot 10^8$ & $3.9\cdot 10^9$ & $2.6\cdot 10^8$ \\[1pt] \midrule
shapes per second & $3\cdot 10^5$ & $7\cdot 10^4$ & $2\cdot 10^4$ \\[1pt] 
comp. time per run, $T/I$ [s] & $4.8 \cdot 10^{-2}$ & $1.2 \cdot 10^0$ & $1.2 \cdot 10^1$ \\[1pt]
evaluated shapes per run & $1.4\cdot 10^4$ & $7.8\cdot 10^4$ & $2.6\cdot 10^5$ \\[1pt]
\midrule
$Q_\T{min} / Q_\T{lb}^\T{TM}$ & $1.18$ & $1.12$ & $1.11$ \\[1pt] \bottomrule
\multicolumn{4}{r}{Computer: CPU Threadripper 1950 ($3.4$\,GHz), $128$\,GB RAM.}\\[1pt]
\end{tabular} 
\label{Tab:compTime1}
\end{table}

\begin{figure}
	\centering
	\includegraphics[width=\figwidth cm]{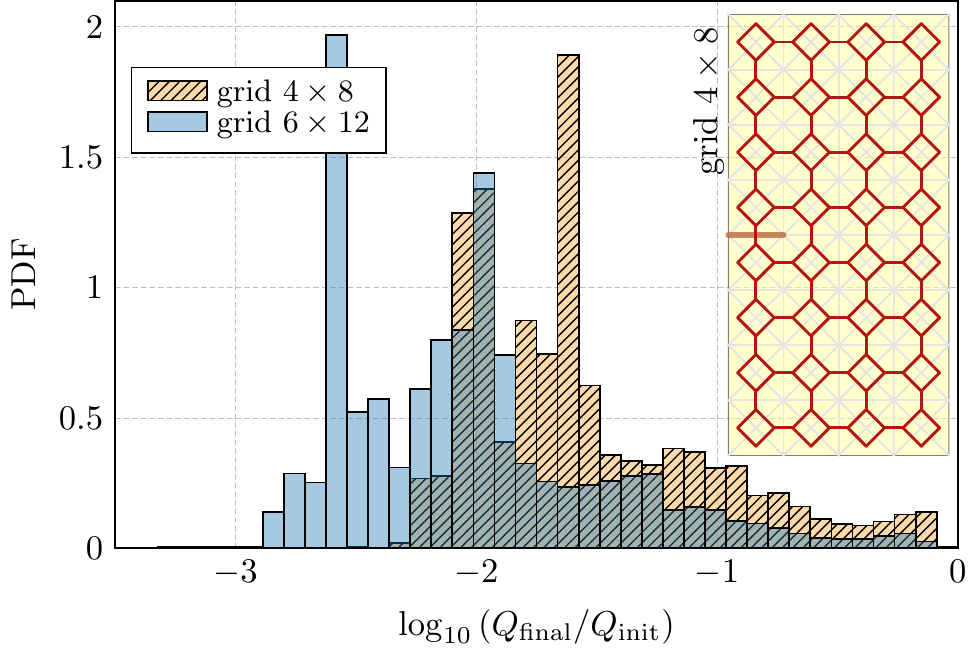}
	\caption{Probability density function (PDF) representing relative improvements of the criterion function (Q-factor) between the locally optimal shape and the initial sample selected by the Monte Carlo algorithm ($5\cdot 10^4$ iterations were used). It is seen that the criterion function value is, on average, minimized to approximately $0.06$ and to approximately $0.03$ of the initial guess for $4\times 8$ and $6\times 12$ grids, respectively.}
	\label{fig:fig3NB}
\end{figure}

\begin{figure}
	\centering
	\includegraphics[width=\figwidth cm]{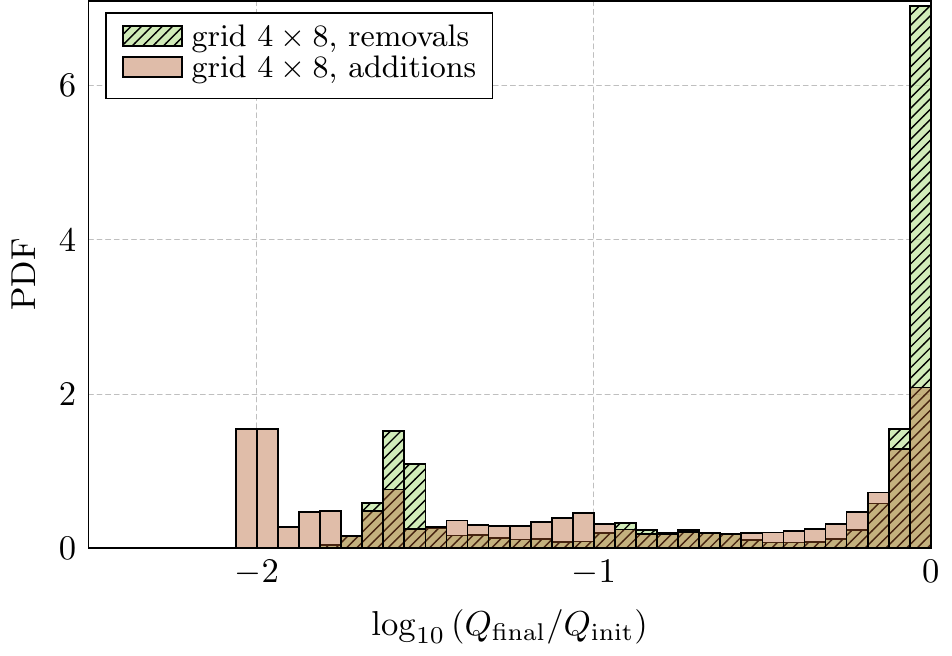}
	\caption{Probability density function (PDF) representing the relative improvements of the criterion function (Q-factor) between the locally optimal shape and the initial sample selected by the Monte Carlo algorithm ($5\cdot 10^4$ iterations were used). Either sole basis function removals~\eqref{eq:topoRemI} or sole basis function additions~\eqref{eq:topoAddI} were allowed.}
	\label{fig:fig3NA}
\end{figure}

It can be seen in Fig.~\ref{fig:fig3NB} that the greedy search, which is a local algorithm, is capable of improving the performance in Q-factor to a mean value of $0.064 Q_\T{init}$ and $0.028 Q_\T{init}$ for $4\times 8$ and $6\times 12$ grids, respectively. As expected, the improvements are more pronounced when working with a finer grid (non-hatched bars). As confirmed in Fig.~\ref{fig:fig3NA}, this is only possible if both removal and addition techniques are involved. In particular, the removal alone performs poorly when starting from a random seed. Figure~\ref{fig:fig4N} shows the statistics of resemblance between the initial and final samples and reveals that, on average, $\kappa \approx 0.43$ of all \ac{dof} were modified. The final result, showing the probability density function (PDF) and the cumulative density function (CDF) of a normalized Q-factor, is presented in Fig.~\ref{fig:fig5N}. Interestingly, the most probable Q-factor is close to its lower bound and is reachable from many starting positions.

\begin{figure}
\centering
\includegraphics[width=\figwidth cm]{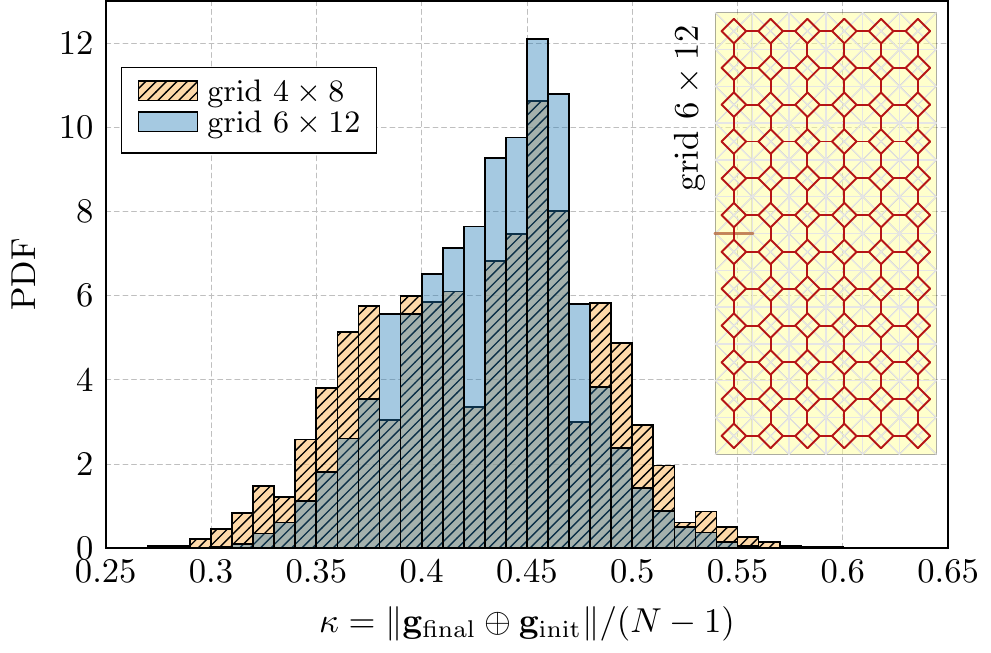}
\caption{Probability density function (PDF) representing the number of local steps, \ie{}, the number of iterations of the greedy algorithm performed to find the local minimum.}
\label{fig:fig4N}
\end{figure}

\begin{figure}
\centering
\includegraphics[width=\figwidth cm]{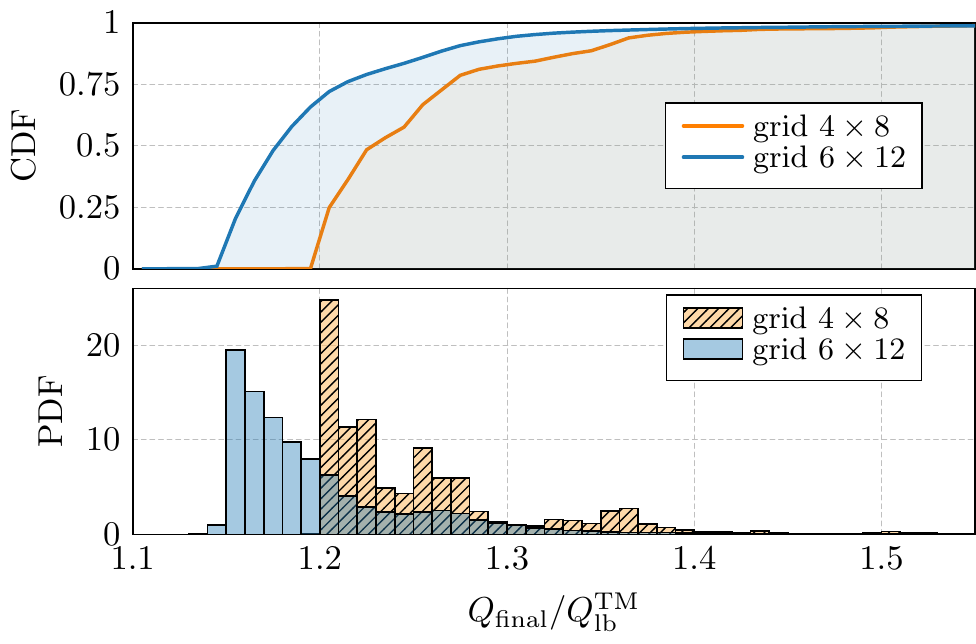}
\caption{Probability density function (PDF) and cumulative density function (CDF) representing local minima found by the greedy algorithm starting from positions generated by a Monte Carlo algorithm.}
\label{fig:fig5N}
\end{figure}

\section{Conclusion}
\label{sec:concl}

An inversion-free method was introduced for evaluation of the smallest perturbation within the \ac{MoM}. It makes it possible to preserve the binary nature of the shape synthesis problem, furthermore improving the convergence rate and robustness of the optimization method. The reconstruction of the structure was derived using the Sherman-Morrison-Woodbury identity.

A greedy algorithm used on topology sensitivity was employed to demonstrate the capability to gather millions of evaluated shapes per minute. To this end, the presented method can be utilized as a local step in global optimization schemes. When randomly restarted, it can also serve as a data mining tool or as a building block for machine learning techniques aimed at shape/pattern synthesis.

The letter also stimulates further development. It is inevitable that a study of the dependence of the method on the number of unknowns and the type of mesh grid will be required. Another important question to be discussed concerns the multi-objective formulation and the proposal of a hybrid optimization algorithm based on a combination of a heuristic approach and topology sensitivity.

\bibliographystyle{IEEEtran}
\bibliography{references}

\end{document}